\documentclass[conf ]{new-aiaa}
\usepackage[utf8]{inputenc}
\usepackage{textcomp}

\usepackage{graphicx}
\usepackage{amsmath}
\usepackage{color}
\usepackage[version=4]{mhchem}
\usepackage{siunitx}
\usepackage{longtable,tabularx}
\title{Total reflection of two guided waves for embedded trapped modes}

\author{Xiwen Dai \footnote{xiwen.dai@sjtu.edu.cn.}}
\affil{School of Mechanical Engineering, Shanghai Jiao Tong University, Shanghai 200240, China}

\begin{document}

\maketitle

\begin{abstract}
To investigate the mechanism of wave trapping, acoustic embedded trapped modes associated with two-resonant-mode interference in two-dimensional duct--cavity structures are calculated by the feedback-loop closure principle, which allows us to analyse the travelling modes that construct the trapped modes.
The exact two coexisting resonant modes that underpin an embedded trapped mode in a cavity open to two semi-infinite ducts are numerically demonstrated.
At the interface between the cavity segment and a duct, total reflection can occur for a particular combination of two propagative guided waves in the cavity. With a particular cavity length and a particular frequency, total reflection also happens for the two reflected waves at the opposite end of the cavity. In this way, the two coexisting standing waves underpin a trapped mode. 
It is found that the two standing waves are not two closed-cavity modes.
Thus, this work presents a new understanding of such embedded trapped modes as a product of total reflection of two guided waves at the interface between two waveguides, rather than interference between two eigenmodes of a closed cavity.
For quasi-trapped modes, the Fano scattering phenomenon owing to the effects of two acoustic channels is also shown.
\end{abstract}

\section{Introduction}
The global instability and the consequent self-sustained oscillations of a cavity flow become much stronger when the frequency of flow--acoustic resonance approaches the acoustic resonance frequency of the cavity \citep{Yamouni2013}. 
This interaction happens not only in cavities \citep{East1966,Tam1978,Rockwell1978,Yamouni2013} or duct side-branches \citep{Bruggeman1991,Ziada1999}, but also over cascades of flat plates \citep{Parker1966,Parker1967,Koch1983} or plates in a flow duct \citep{Cumpsty1971,Stoneman1988} and in aeroengines \citep{Cooper2000,Hellmich2008}. Therefore, acoustic resonance in an open system, i.e. acoustic trapped mode, is a problem of interest in many flow--acoustic coupling problems. 

Trapped modes, i.e. complete wave localization in open systems, are found in such as optical, water and acoustic waves \citep{Hsu2016}. 
A trapped mode residing inside the continuous spectrum of propagative waves is referred to as an embedded trapped mode or a bound state in the continuum (BIC). The most common mechanism of an embedded trapped mode is based on the symmetry of the system, i.e. symmetric mismatch between the trapped mode and the continuous-spectrum modes \citep{Evans1991,Evans1994,Evans1997}. However, the symmetry argument is violated by a second type of embedded trapped mode associated with two-resonant-mode interference. It is often named after Friedrich and Wintgen \citep {Friedrich1985} who proposed that a BIC can result from two interfering resonances. A third type is due to Fabry--P{\'e}rot interference between two resonant reflectors acting as `mirrors' \citep{McIver1996,Kuznetsov2001}. More discussions of the mechanisms and theoretical studies of embedded trapped modes can be found in Refs. \citep{Linton2007,Pagneux2013,Hsu2016}, and references therein.

Trapped modes can be numerically solved from the global eigenvalue problem in a finite domain with absorbing boundary conditions \citep{Hein2004,Koch2005,Duan2007,Hein2007,Hein2010,Hein2012}. This approach provides all the spectra and eigenfunctions of trapped and quasi-trapped modes for any complex structures. All the three types of embedded trapped modes in acoustics in duct--cavity systems have been numerically demonstrated in Ref. \cite{Hein2012}. Recently, the acoustic coupled mode approach, i.e. the non-Hermitian Hamiltonian approach known in quantum mechanics, has been introduced for the calculation of trapped modes in acoustics \citep{Lyapina2015,Xiong2016,Xiong2016b}. Trapped modes were calculated by projecting them to closed-cavity modes. With this approach, Ref. \cite{Lyapina2015} has made an effort to numerically demonstrate the two resonant modes for the second type embedded trapped mode. 
Since the building bricks of trapped modes were closed-cavity modes and the resemblance in the pressure fields between a trapped mode and the superposition of two closed-cavity modes was observed,  a `two-mode approximation', that is, using two closed-cavity modes to approximate a trapped mode, has been proposed. However, it has also been revealed in that paper that to accurately represent the pressure field of a trapped mode, one needs more than two closed-cavity modes, and the embedded trapped modes do not occur at the crossing points of the eigenfrequencies of closed-cavity modes as the cavity length is varied. The shift of the trapped mode frequencies from the crossing points has been shown to be small in many cases, but rather large in some cases \citep{Hein2012,Lyapina2015}. The inaccuracy of the `two-mode approximation' was attributed to the coupling with evanescent modes in the ducts \cite{Lyapina2015}.

\begin{figure}
	\begin{center}
		\includegraphics[width=12cm]{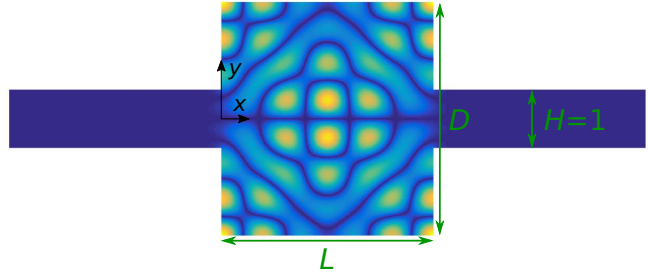}
	\end{center}	
	\caption{\label{fig_1} {(Colour online) Iso-colour plot of $|p|$ of an embedded trapped mode in a 2-D symmetric duct--cavity structure, i.e. a cavity open to two semi-infinite ducts, with frequency $\omega_t=6.5044$, $D=4$, and $L=3.6331$.}}
\end{figure}

In this paper, acoustic embedded trapped modes associated with two-resonant-mode interference in two-dimensional duct--cavity structures, as sketched in Fig \ref{fig_1}, are calculated by the feedback-loop closure principle \citep{Landau1981}, which allows us to construct trapped modes with the travelling duct modes. It has been shown that the combination of travelling mode and global mode analyses provides additional insight into global mode or flow--acoustic resonance in flow \citep{Dai2020}. This work is also stimulated by the understanding that the underlying physics of trapped or quasi-trapped modes is wave reflection at effective boundaries, which are the interfaces between the cavity and the ducts in the present cases.
In contrast to Ref. \cite{Lyapina2015}, the present study demonstrates the exact two coexisting resonant modes that underpin an embedded trapped mode. Those two coexisting resonant modes are found to be two standing waves in the cavity, but they are not two closed-cavity modes, which explains why one needs more than two closed-cavity modes to represent a trapped mode and also the frequency shift of a trapped mode from the crossing point of the eigenfrequencies of two closed-cavity modes. 
It is noted that the two standing waves discussed here are decoupled with the outgoing propagative modes but coupled with the outgoing evanescent modes in the ducts, which are well-known facts for embedded trapped modes in such an open system.
Thus, Ref. \cite{Lyapina2015} has not found the exact two coexisting resonant modes that underpin an embedded trapped mode not because the trapped mode is coupled with the outgoing evanescent modes in the ducts, but because at the interfaces between the cavity and the ducts the \emph{a priori} acoustically hard-wall condition was enforced for each component mode, i.e. a closed-cavity mode \cite{Lyapina2015}. Such constraint for the component resonant modes is released in the present approach.
The present finding leads to a new understanding of the mechanism of the embedded trapped modes considered, that is, a product of total reflection for a particular combination of two propagative guided waves at the interface between two waveguides, rather than interference between two eigenmodes of a closed cavity \citep{Lyapina2015}. And, total reflection of multiple travelling duct modes is understood as a result of linear cancellation in transmission to the propagative channel(s) of the adjacent waveguide.

The numerical model is described in Sec. \ref{sec2}. 
In Sec. \ref{sec3a}, results demonstrate the wave trapping mechanism for the embedded trapped modes considered, i.e. total reflection for a particular combination of two propagative guided waves. 
Some notes on total reflection of multiple guided waves are given in Sec. \ref{sec3b}. It is highlighted that neither a resonator nor a cavity is relevant to such total reflection.
The Fano scattering \citep{Fano1961,Miroshnichenko2010,Lukyanchuk2010} of a quasi-trapped mode is then investigated in Sec. \ref{sec3c}. It is shown that the transmission can be accurately described by the contributions of two acoustic channels.

\section{Numerical model}
\label{sec2}

\subsection{Calculation of modal scattering}
Without flow, the governing equation of acoustic disturbances is a wave equation. For a 2-D problem, the normalized equation is written,
\begin{equation}
\frac{\partial ^2 p}{\partial t ^2} -\left( \frac{\partial^2p}{\partial x^2}+\frac{\partial^2 p}{\partial y^2}\right)=0, \label{eq31}
\end{equation}
where $p$ is pressure, $t$ is time, $x$ and $y$ are the coordinates along and perpendicular to the duct axis.
The fluctuation is sought in the form of transverse modes,
\begin{equation}
p = P(y)\exp(-\mathrm{i} k x)\exp(\mathrm{i} \omega t ), \label{eq32}
\end{equation}
where $\mathrm{i}^2=-1$, $k$ is wavenumber, and $\omega$ is the angular frequency.
Inserting Eq. (\ref{eq32}) into Eq. (\ref{eq31}) leads to:
\begin{equation}
 k^2 P- \omega ^2 P-\frac{\mathrm{d} ^2 P}{\mathrm{d} y^2}=0. \label{eq33}
\end{equation}

The duct--cavity structure shown in Fig. \ref{fig_1} is divided into three zones: the left duct (zone I),  the cavity segment (zone II), and the right duct (zone III). In each zone, duct modes are solved by discretizing Eq. (\ref{eq33}) in the $y$-direction \citep{Kooijman2008,Dai2018}, taking $N_1$ equally spaced points in the ducts, $N_2$ equally spaced points in the cavity segment. The spacing between interior points in all zones is $\Delta h =H/N_1= D/N_2$, and the first and last points are taken $\Delta h/2$ from the solid walls. 
The second-order centered finite difference method is used to solve the eigenvalue problems for the transverse modes (using the \texttt{eig} function of MATLAB).
In the ducts, $2N_1$ transverse modes are found: $N_1$ modes propagating or decaying in the $\pm x$ directions. 
In the cavity segment, $2N_2$ transverse modes are solved: $N_2$ modes propagating or decaying in the $\pm x$ directions. 

The transverse modes in different zones are then matched using the continuity of pressure $p$ and $\partial p/ \partial x$ at the interfaces between zones, and $\partial p/ \partial x=0$ on the vertical walls inside the cavity. The match leads to the scattering matrix of the cavity, which links outgoing waves to incoming waves,
\begin{eqnarray}
\left(\begin{array}{c}
\mathbf{C}_3^+ \\ \mathbf{C}_1^- \\
\end{array}\right)= 
\mathbf{S}\left(\begin{array}{c} \mathbf{C}_1^+ \\ \mathbf{C}_3^-\\
\end{array}\right), \label{eq34}
\end{eqnarray}
where vectors $\mathbf{C}^{\pm}_1$ (respectively $\mathbf{C}^{\pm}_3$) contain the transverse mode coefficients for $x=0$ (respectively $x=L$) for waves travelling in the $+x$ (respectively $-x$) direction
\begin{eqnarray}
\mathbf{S}= 
\left(\begin{array}{c c} \mathbf{T}^+ & \mathbf{R}^-  \\ \mathbf{R}^+ & \mathbf{T}^-\\
\end{array}\right), \label{eq35}
\end{eqnarray}
where  $\mathbf{T}^+$ $(N_1\times N_1)$, $\mathbf{R}^+$ $(N_1\times N_1)$, $\mathbf{T}^-$ $(N_1\times N_1)$ and $\mathbf{R}^-$ $(N_1\times N_1)$ are transmission and reflection matrices in the $\pm x$ directions. 

\subsection{Calculation of trapped modes}

The acoustic resonances in the duct--cavity structure shown in Fig. \ref{fig_1} arise from wave reflection at the two ends of the cavity segment. Each trapped or quasi-trapped mode is assembled by the travelling transverse modes that replicate themselves after a feedback loop in the cavity segment at a real-valued or complex frequency.
For a general idea of the feedback-loop closure principle in investigating resonances and global instabilities in flow and fluid-structure systems, the reader is referred to Refs. \cite{Landau1981,Gallaire2004,Alvarez2004,Doare2006,Tuerke2015,Jordan2018}.
For the present calculaitons, the reader is also referred to Ref. \cite{Dai2020}, where the method was used to investigate flow--acoustic resonance in a cavity covered by a perforated plate.

We define a multimodal feedback-loop matrix, $\mathbf{M}_{fl} =  \mathbf{R}_{l}\mathbf{P}_{l}\mathbf{R}_r\mathbf{P}_{r}$,
where $\mathbf{R}_{l}$ $(N_2\times N_2)$ and $\mathbf{R}_r$ $(N_2 \times N_2)$ are reflection matrices at the left and right ends of the cavity segment respectively. $\mathbf{R}_{l}$ (respectively $\mathbf{R}_r$) links $k^+$ and $k^-$ travelling modes in the cavity segment at the left (respectively right) end of the cavity, and they are calculated by matching the transverse modes at the two ends of the cavity segment, with no incoming waves in the two semi-infinite ducts being present. $\mathbf{P}_{l}$ and $\mathbf{P}_{r}$ ($N_2 \times N_2$ diagonal matrices with on the diagonal the values of $\exp(\mathrm{i} k^-_n L)$ and $\exp(-\mathrm{i} k^+_n L)$ respectively, where $k^{\mp}_n$ are the wavenumbers of the $n^{th}$ left- and right-travelling modes) stand for wave propagation inside the cavity segment in the $\mp x$ directions. The loop closure principle means that one of the eigenvalues of $\mathbf{M}_{fl}$ is unity at the complex frequency of a resonance: $\mathbf{M}_{fl} \mathbf{C}_{fl}= k_{fl} \mathbf{C}_{fl}$ with $k_{fl} =1$, 
where $k_{fl}$ and $\mathbf{C}_{fl}$ are the unity eigenvalue and the corresponding eigenvector. 
The eigenvector contains the coefficients of the transverse modes that lead to the field distribution of the trapped mode, which is an eigenfunction of the global eigenvalue problem \citep{Hein2012}.
In some situations, these two approaches can be proved to be equivalent \citep{Gallaire2004}.

In the calculations, the real-valued frequency ($\mathrm{Im}(\omega_t)=0$) and the cavity length $L$ are optimized for a trapped mode (using the \texttt{fminsearch} function of MATLAB), for that one of the eigenvalues of $\mathbf{M}_{fl}$ equals to unity.
When the cavity length is perturbed from the value for a trapped mode, $\mathrm{Re}(\omega_{qt})$ and $\mathrm{Im}(\omega_{qt})$ are optimized for a quasi-trapped mode. The iteration stops when the error between the target eigenvalue and unity is less than $E_o=10^{-12}$. 
$N_1=400$ is used for the present calculations and a convergence study of the numerical calculations is given in Appendix A. 

\section{Results}
\label{sec3}

\subsection{Embedded trapped modes owing to total reflection of two guided waves}
\label{sec3a}

\begin{figure}
	\begin{center}
		\includegraphics[width=13cm]{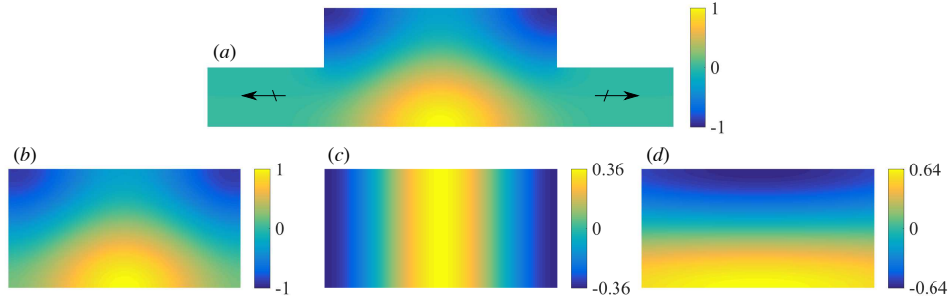}
	\end{center}	
	\caption{\label{fig_2} {(Colour online) Embedded trapped mode in a 2-D cavity open to two semi-infinite ducts with $\omega_t=1.5957$, $D=2$, and $L=3.9375$. Iso-colour plots are $\mathrm{Re}(p)$ of (a) the trapped mode, (b) superposition of two standing waves in the cavity region, (c) one standing wave with amplitude 0.3554, and (d) the other standing wave with amplitude 0.6446. The arrows denote the decoupled outgoing propagative modes, same in Figs. \ref{fig_4}, \ref{fig_6} and  \ref{fig_13}.}}
\end{figure}

\begin{figure}
	\begin{center}
		\includegraphics[width=12cm]{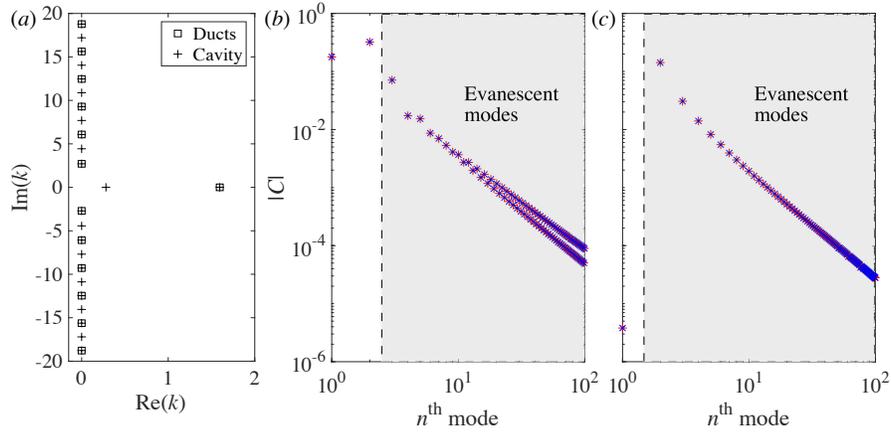}
	\end{center}	
	\caption{\label{fig_3} {(Colour online) Wavenumbers (a) and mode coefficients (b,c) of the transverse modes associated with the trapped mode in Fig. \ref{fig_2}. In (b), \textcolor{red}{$+$} (respectively \textcolor{blue}{$\times$}) denote coefficients of $k^+$ (respectively $k^-$) modes at $x=0$ (respectively $x=L$) in the cavity. In (c), \textcolor{red}{$+$} (respectively \textcolor{blue}{$\times$}) denote coefficients of $k^+$ (respectively $k^-$) modes at $x=L$ (respectively $x=0$) in the right (respectively left) duct.}}
\end{figure}

\begin{figure}
	\begin{center}
		\includegraphics[width=13cm]{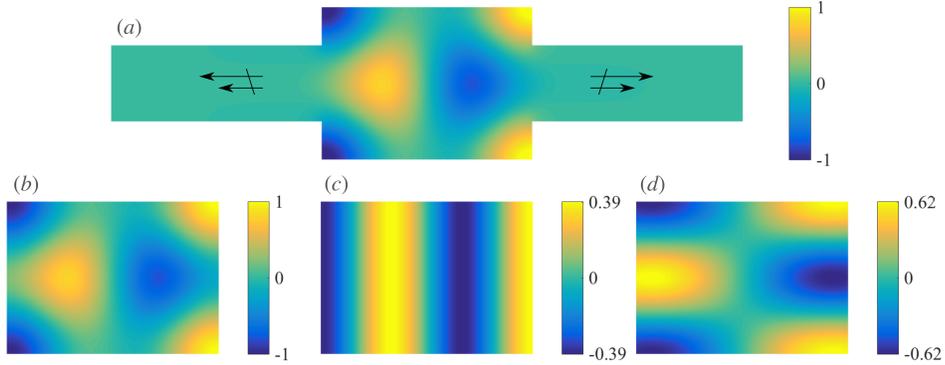}
	\end{center}	
	\caption{\label{fig_4} {(Colour online) Embedded trapped mode in a 2-D cavity open to two semi-infinite ducts with $\omega_t=3.4069$, $D=2$, and $L=2.7664$. Iso-colour plots are $\mathrm{Re}(p)$ of (a) the trapped mode, (b) superposition of two standing waves in the cavity region, (c) one standing wave with amplitude 0.3916, and (d) the other standing wave with amplitude 0.6244.}}
\end{figure}

\begin{figure}
	\begin{center}
		\includegraphics[width=12cm]{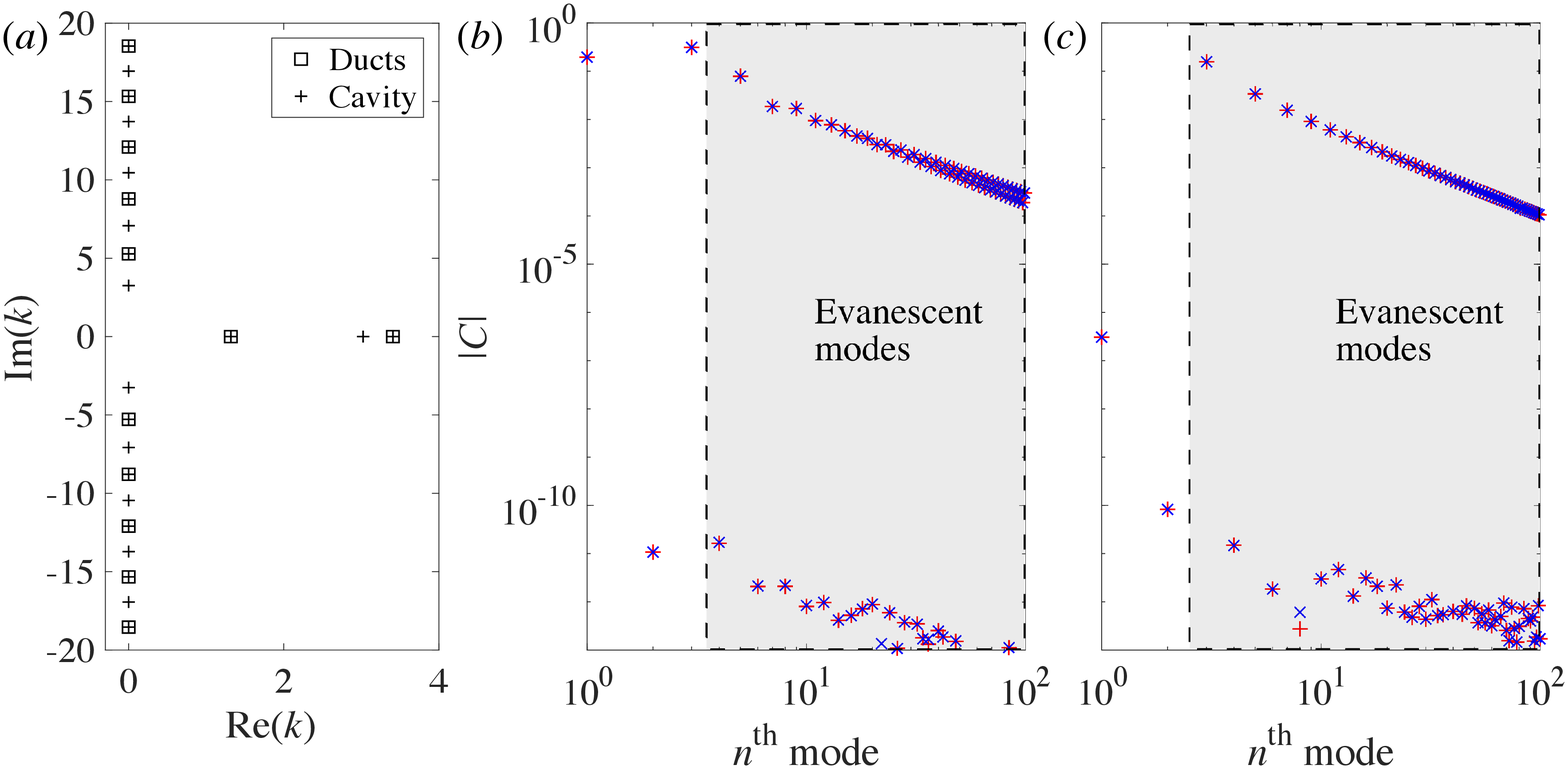}
	\end{center}	
	\caption{\label{fig_5} {(Colour online) Wavenumbers (a) and mode coefficients (b,c) of the transverse modes associated with the trapped mode in Fig. \ref{fig_4}. In (b), \textcolor{red}{$+$} (respectively \textcolor{blue}{$\times$}) denote coefficients of $k^+$ (respectively $k^-$) modes at $x=0$ (respectively $x=L$) in the cavity. In (c), \textcolor{red}{$+$} (respectively \textcolor{blue}{$\times$}) denote coefficients of $k^+$ (respectively $k^-$) modes at $x=L$ (respectively $x=0$) in the right (respectively left) duct.}}
\end{figure}

\begin{figure}
	\begin{center}
		\includegraphics[width=13cm]{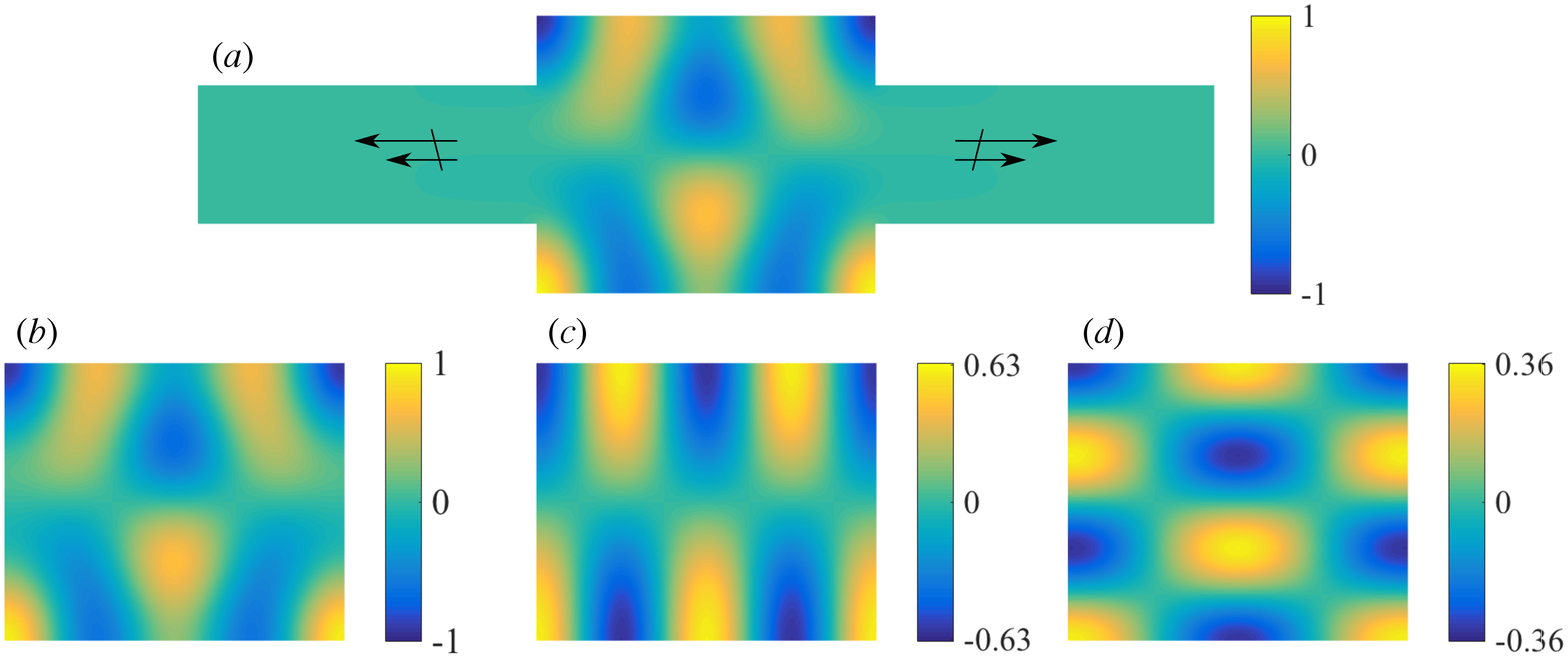}
	\end{center}	
	\caption{\label{fig_6} {(Colour online) Embedded trapped mode in a 2-D cavity open to two semi-infinite ducts with $\omega_t=5.3971$, $D=2$, and $L=2.4356$. Iso-colour plots are $\mathrm{Re}(p)$ of (a) the trapped mode, (b) superposition of two standing waves in the cavity region, (c) one standing wave with amplitude 0.6372, and (d) the other standing wave with amplitude 0.3633.}}
\end{figure}

\begin{figure}
	\begin{center}
		\includegraphics[width=12cm]{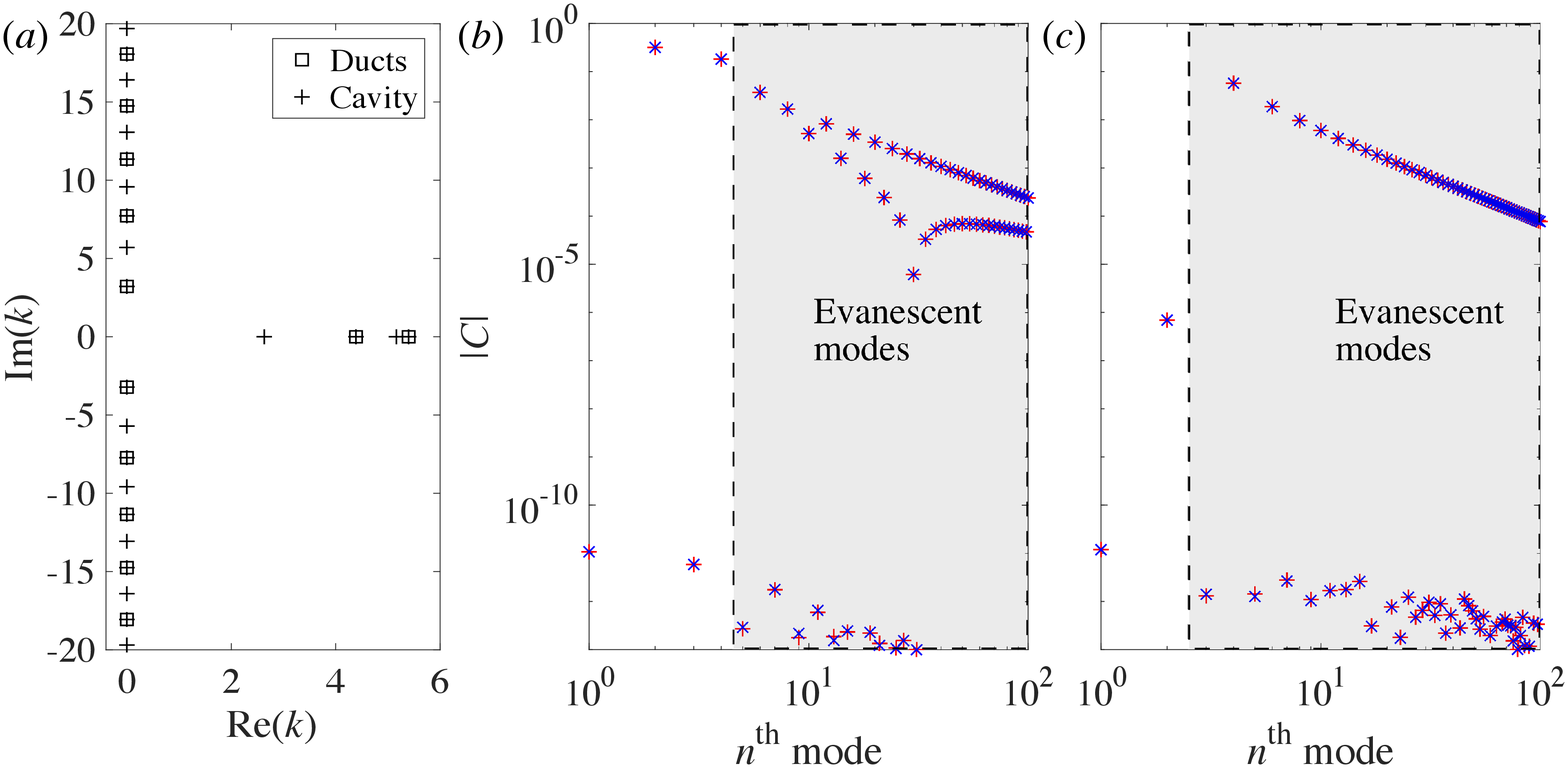}
	\end{center}	
	\caption{\label{fig_7} {(Colour online) Wavenumbers (a) and mode coefficients (b,c) of the transverse modes associated with the trapped mode in Fig. \ref{fig_6}. In (b), \textcolor{red}{$+$} (respectively \textcolor{blue}{$\times$}) denote coefficients of $k^+$ (respectively $k^-$) modes at $x=0$ (respectively $x=L$) in the cavity. In (c), \textcolor{red}{$+$} (respectively \textcolor{blue}{$\times$}) denote coefficients of $k^+$ (respectively $k^-$) modes at $x=L$ (respectively $x=0$) in the right (respectively left) duct.}}
\end{figure}

We first calculate three embedded trapped modes considered in \cite{Xiong2016}: case 1 (Figs. \ref{fig_2} and \ref{fig_3}) and in \cite{Lyapina2015}: case 2 (Figs. \ref{fig_4} and \ref{fig_5}) and case 3 (Figs. \ref{fig_6} and \ref{fig_7}).
In the pressure fields shown in Figs. \ref{fig_2}, \ref{fig_4}, and \ref{fig_6}, $\mathrm{Re}(p)$ is normalized by the maximum of $|p|$ in the subfigure (b) of each figure. 
The corresponding wavenumbers and mode coefficients of the transverse modes are shown in Figs. \ref{fig_3}, \ref{fig_5}, and \ref{fig_7}, where the coefficients are presented after the mode profiles being normalized: $|P_n(y)|=1$.

\begin{figure}
	\begin{center}
		\includegraphics[width=12cm]{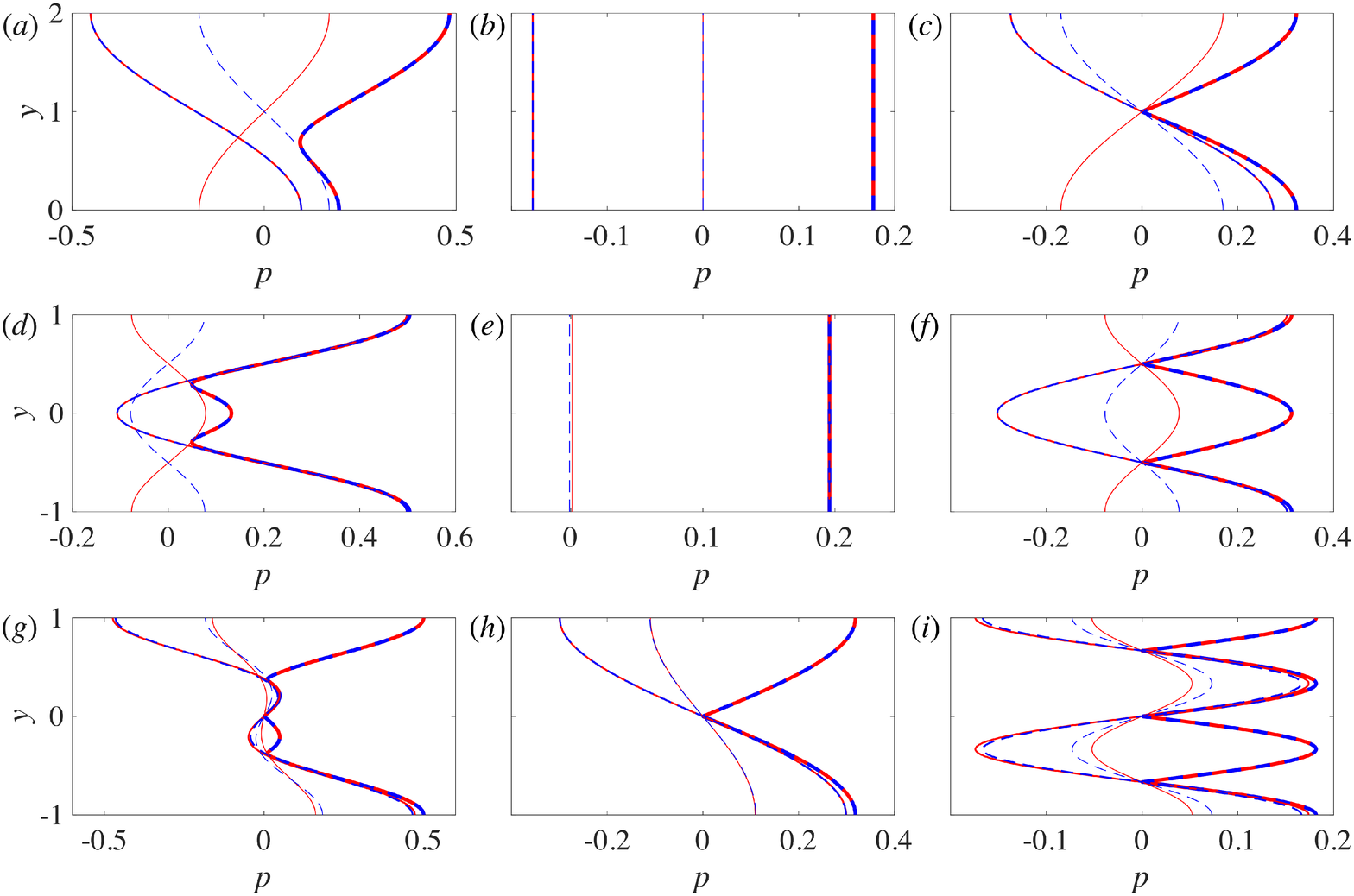}
	\end{center}	
	\caption{\label{fig_8} {(Colour online) Total reflection of two propagative guided waves at the interface between the cavity and a duct. (a-c) is for the trapped mode in Fig. \ref{fig_2} (note that the $x$ coordinate axis is on the bottom wall), (d-f) is for Fig. \ref{fig_4}, and (g-i) is for Fig. \ref{fig_6}. (a,d,g) is for $k1+k2$, (b,e,h) is for $k1$ and (c,f,i) is for $k2$. Red solid lines denote the incident wave(s), blue dashed lines denote the reflected wave(s). Thin lines: $\mathrm{Re}(p)$, lines with medium thickness: $\mathrm{Im}(p)$, thick lines: $|p|$.}}
\end{figure}

In case 1, one transverse mode is propagative in the $\pm x$ directions in the ducts and the outgoing plane waves have a very small amplitude ($<10^{-5}$), as shown in Fig. \ref{fig_3}(c), indicating that an embedded trapped mode is solved numerically.
Two transverse modes in the cavity segment are propagative in the $\pm x$ directions, $k^+_n$ mode and $k^-_n$ mode (where $n=1,2$) have the same amplitude, as shown in Fig. \ref{fig_3}(b). Thus, two standing waves are formed in the cavity segment, as shown in Fig. \ref{fig_2}(c,d). The superposition of these two standing waves is shown in Fig. \ref{fig_2}(b). From the pressure fields of the superposition of the two standing waves and of the trapped mode, we cannot find any visual differences. 
In cases 2 and 3, two transverse modes are propagative in the $\pm x$ directions in the ducts and the outgoing propagative waves have very small amplitudes ($<10^{-6}$), as shown in Figs. \ref{fig_5}(c) and \ref{fig_7}(c), denoting embedded trapped modes numerically solved. In the cavity segment, three (respectively four) transverse modes are propagative in the $\pm x$ directions in case 2 (respectively in case 3). However, $k^{\pm}_2$ modes in case 2 and $k^{\pm}_{1,3}$ modes in case 3 have vanishingly small amplitudes ($<10^{-10}$). The resemblance between a trapped mode and the superposition of two standing waves is also found in Figs. \ref{fig_4} and \ref{fig_6}.

Comparison of the pressure fields suggests that an embedded trapped mode of this type might be represented by two standing waves. We are also aware of the small differences near the two ends of the cavity caused by evanescent modes in the cavity segment. 
The following will show that the mechanism of the trapped modes of the present type can be accurately described by the coexistence of two standing waves (or resonant modes), and the effects of those evanescent modes  in the cavity can be neglected without losing accuracy.

Figures \ref{fig_2}(d), \ref{fig_4}(d), and \ref{fig_6}(d) show that, in each case, one of the two standing waves is not a closed-cavity mode, which leads us to examine the wave reflection at the interfaces between the cavity and the ducts. Taking the interface ($x=L$) between the cavity and the right duct for example, the highest amplitude of $k^+$ evanescent waves in the cavity at $x=L$ is $4.47 \times 10^{-6}$ in case 1 ($9.46 \times 10^{-8}$ in case 2 and $9.21 \times 10^{-8}$ in case 3), thus the only incident waves onto this interface are the two $k^+$ propagative waves, named as $k1$ and $k2$ (Note that the other $k^+$ propagative waves in cases 2 and 3 have vanishingly small amplitudes). 
The pressure $p(y)$ at $x=L$ associated with the two incident waves ($k1$ and $k2$) and the two reflected waves are plotted in Fig. \ref{fig_8}, total reflection ($|p^+|=|p^-|$) for both $k1$ and $k2$ is observed. It is noted that the total reflections of $k1$ and $k2$ are coupled, since the two-by-two reflection matrix is not a diagonal matrix (e.g. $\mathbf{R}(2\times 2)=[0.7215 + 0.0936\mathrm{i}, 0.1578 + 0.0367\mathrm{i}; 1.7918 + 0.4174\mathrm{i},-0.2717 + 0.1793\mathrm{i}]$ in case 1). This total reflection of two incident guided waves is a local wave scattering phenomenon at the interface between two waveguides with different cross-areas, meaning that the cavity length does not have a relevant effect. Total reflection requirements at both ends select the cavity length for a trapped mode. 
It is also shown that the total reflection for $k2$ is not a hard-wall reflection ($p^+=p^-$), which explains the difference between the corresponding standing wave and a closed-cavity mode. For $k1$ the results indicate a hard-wall reflection.
Nevertheless, it has been shown in Ref. \cite{Lyapina2015} that, in an axisymmetric duct-cavity structure, the frequency of an embedded trapped mode can deviate from the frequencies of all closed-cavity modes, which means that the total reflection for both $k1$ and $k2$ could be different from a hard-wall reflection in some cases.
Since the incident waves for total reflection on the effective boundaries of an embedded trapped mode can be accurately represented by two waves, the mechanism of such a trapped mode can be accurately described by the coexistence of two standing waves. 
It is also noted that the two coexisting standing waves are certainly coupled with the outgoing evanescent modes in the ducts, as indicated by the non-vanishing amplitudes of the evanescent modes in Figs. \ref{fig_3}(c), \ref{fig_5}(c) and \ref{fig_7}(c). 

\begin{figure}
	\begin{center}
		\includegraphics[width=10cm]{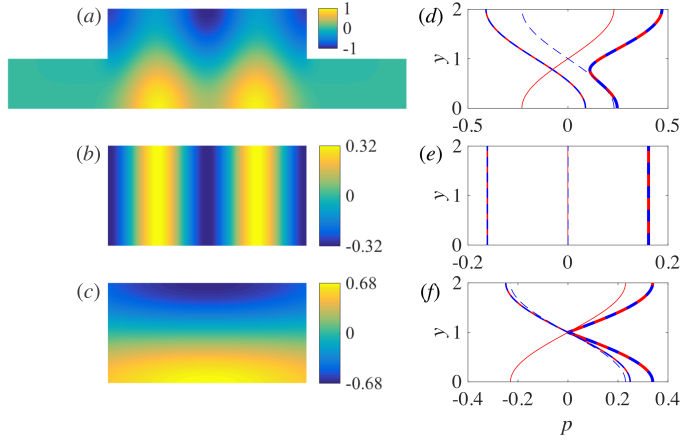}
	\end{center}	
	\caption{\label{fig_9} {(Colour online) Embedded trapped mode in a 2-D cavity open to two semi-infinite ducts with $\omega_t=1.5820$, $D=2$, and $L=7.9433$. Iso-colour plots are $\mathrm{Re}(p)$ of (a) the trapped mode, (b) one standing wave with amplitude 0.3223, and (c) the other standing wave with amplitude 0.6777. Note that in (a-c) the plots have been scaled by a factor of 1/2 in the $x$ direction. (d-f) total reflection of guided wave(s) at the right end of the cavity. Red solid lines denote the incident wave(s), blue dashed lines denote the reflected wave(s). Thin lines: $\mathrm{Re}(p)$, lines with medium thickness: $\mathrm{Im}(p)$, thick lines: $|p|$.}}
\end{figure}

\begin{figure}
	\begin{center}
		\includegraphics[width=11cm]{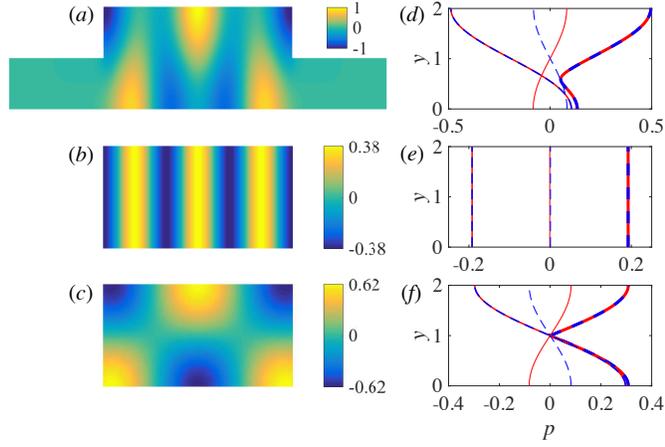}
	\end{center}	
	\caption{\label{fig_10} {(Colour online) Embedded trapped mode in a 2-D cavity open to two semi-infinite ducts with $\omega_t=1.6853$, $D=2$, and $L=11.1850$. Iso-colour plots are $\mathrm{Re}(p)$ of (a) the trapped mode, (b) one standing wave with amplitude 0.3842, and (c) the other standing wave with amplitude 0.6158. Note that in (a-c) the plots have been scaled by a factor of 1/3 in the $x$ direction. (d-f) total reflection of guided wave(s) at the right end of the cavity. For the descriptions of the lines in (d-f), see Fig. \ref{fig_9}.}}
\end{figure}

Each case above shows one value of $L$ selected by the total reflection requirements for $k1$ and $k2$ at one end and for the two reflected propagative waves at the opposite end of the cavity. Nevertheless, for the same two transverse modes in the $+x$ or $-x$ direction in each case, the total reflection requirements can be fulfilled with multiple values of the cavity length. Figures \ref{fig_9} and \ref{fig_10} show another two embedded trapped modes that are produced by the total reflection of the same two guided waves $k^+_{1,2}$ at $x=L$ and of the two reflected waves $k^-_{1,2}$  at $x=0$ as those in Figs. \ref{fig_2}. It is shown in Figs. \ref{fig_9} and \ref{fig_10} that the total reflection is still a hard-wall reflection for $k^+_1$, and for $k^+_2$ there is still a phase change on wave reflection. Since for a trapped mode the phase changes of both $k^+_1$ and $k^+_2$ modes around a feedback loop should be an integral multiple of $2\pi$, $k^+_1$ requires $L_n=nL_1$ ($L_1$ is the cavity length in Fig. \ref{fig_2}, $L_n$ denotes the possible cavity lengths for trapped modes and $n=1,2,3...$) but $k^+_2$ requires $L_n \neq nL_1$, if the frequency remains the same. This incompatibility explains the frequency shift of the trapped modes presented in Figs. \ref{fig_9} and \ref{fig_10} from that in Fig. \ref{fig_2} and also the shift of the cavity length from $nL_1$. It should be stressed that the shifts are not due to those evanescent modes in the cavity segment, since the the highest amplitude of $k^+$ evanescent waves in the cavity at $x=L$ is $7.34 \times 10^{-11}$ in Fig. \ref{fig_9} ($2.74 \times 10^{-14}$ in Fig. \ref{fig_10}). Owing to the shifts, the trapped modes in longer cavities do not always look like a periodic repeating of that in the shortest cavity, as shown in Fig. \ref{fig_10}.
Trapped modes, owing to the total reflection of the same $k1$ and $k2$ modes as in Fig. \ref{fig_4}, are also calculated for larger $L$: $L=3.4003$ and $\omega_{t}=3.6957$; $L=5.5925$ and $\omega_{t}=3.3705$; $L=7.4189$ and $\omega_{t}=3.8111$; $L=8.4201$ and $\omega_{t}=3.3580$; $L=9.0993$ and $\omega_{t}=3.4526$; $L=10.3268$ and $\omega_{t}=3.9506$. With the frequency being shifted, total reflections still happen at both ends when $L$ deviates a lot from $nL_1$ ($L_1$ is the cavity length in Fig. \ref{fig_4} and $n=1,2,3...$). These trapped modes are either $x$-symmetric or  $x$-antisymmetric.

\begin{figure}
	\begin{center}
		\includegraphics[width=9.5cm]{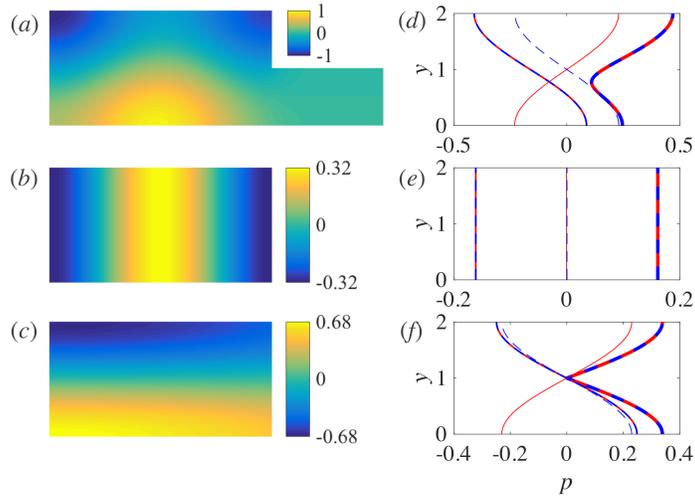}
	\end{center}	
	\caption{\label{fig_11} {(Colour online) Embedded trapped mode in a 2-D cavity open to one semi-infinite duct (a Helmholtz resonator with an infinitely-long neck), with $\omega_t=1.5820$, $D=2$, and $L=3.9717$. Iso-colour plots are $\mathrm{Re}(p)$ of (a) the trapped mode, (b) one standing wave with amplitude 0.3223, and (c) the other standing wave with amplitude 0.6777. (d-f) total reflection of guided wave(s) at the opening end of the cavity. For the descriptions of the lines in (d-f), see Fig. \ref{fig_9}.}}
\end{figure}

In a last case in this subsection, Fig. \ref{fig_11}, the left end of the cavity is closed, which makes the structure $x$-asymmetric. The numerical result also indicates an embedded trapped mode in this Helmholtz resonator. Owing to the asymmetric boundary conditions on the left and right ends of the cavity, the wave reflection at the two ends are different. The total reflection for the $k^+_2$ mode at the opening end of the cavity is not a hard-wall reflection, as shown in Fig. \ref{fig_11}(f), whereas it is a hard-wall reflection for each $k^-$ mode at the left end. The highest amplitudes of $k^+$ evanescent modes in the cavity at $x=0$ and $x=L$ are $1.35 \times 10^{-6}$ and $2.80 \times 10^{-11}$ respectively, whereas the highest amplitudes of $k^-$ evanescent modes at $x=L$ and $x=0$ are $6.46 \times 10^{-2}$ and $1.35 \times 10^{-6}$ respectively. However, $k^+_{1,2}$ modes inside the cavity have the same amplitudes as those two left-travelling guided modes $k^-_{1,2}$, leading to two standing waves as shown in Fig. \ref{fig_11}(b,c).  Since the highest amplitudes of $k^+$ evanescent modes at $x=L$ is vanishingly small ($2.80 \times 10^{-11}$), the incident waves inside the cavity at the opening end can be accurately represented by $k^+_{1,2}$ modes and the embedded trapped mode is generated by the total reflection of $k^+_{1,2}$ modes at the cavity opening. Note that the trapped mode in Fig. \ref{fig_11}(a) is identical to the right half of that in Fig. \ref{fig_9}(a) owing to the $x$-symmetry in Fig. \ref{fig_9}(a).

\subsection{Notes on total reflection of guided waves and trapped modes}
\label{sec3b}

\begin{figure}
	\begin{center}
		\includegraphics[width=13cm]{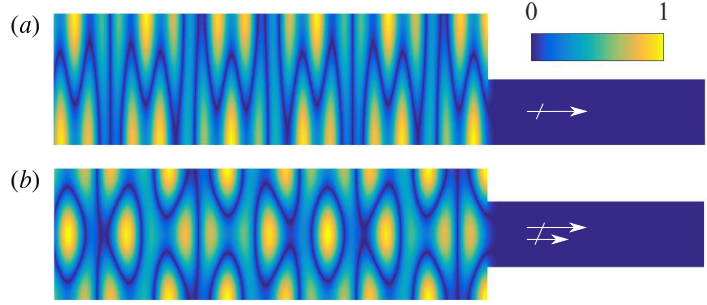}
	\end{center}	
	\caption{\label{fig_12} {(Colour online) Total reflection of two guided waves at the interface between two semi-infinite waveguides. The two incident waves in (a) and (b) are the $k1$ and $k2$ in Figs. \ref{fig_2} and \ref{fig_4} respectively. Note that the iso-colour plots of $|p|$ have been scaled by a factor of 1/9 in (a) and 1/3 in (b) in the $x$ direction. The arrows denote the decoupled $+x$ propagative modes.}}
\end{figure}

Subsection \ref{sec3a} has presented the numerical results of trapped modes, which are solutions of the homogeneous wave equation with outgoing radiation condition, i.e. no incoming waves in the semi-infinite ducts are present. By analysing the component travelling modes of the trapped modes in the open cavity, we have shown that total reflection of two guided waves is the trapping mechanism. In Fig. \ref{fig_12}, the total reflection of $k1$ and $k2$ modes between two semi-infinite waveguides is shown. Note that the $k^+$ evanescent modes of small amplitudes discussed in Figs. \ref{fig_2} and \ref{fig_4} have been excluded from the incident waves in this figure. Neither a resonator nor closed-cavity modes are involved in such a wave scattering.

\begin{figure}
	\begin{center}
		\includegraphics[width=11cm]{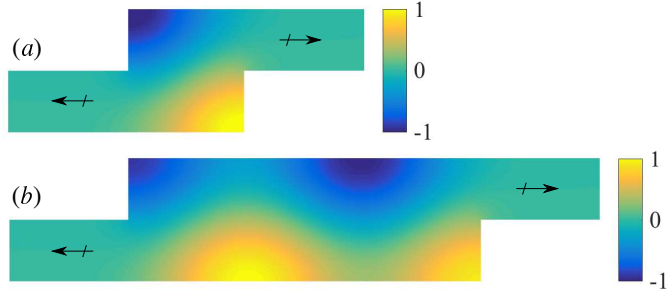}
	\end{center}	
	\caption{\label{fig_13} {(Colour online) Iso-colour plot of $\mathrm{Re}(p)$ of an embedded trapped mode in a 2-D cavity open to two semi-infinite ducts in an $x$-asymmetric manner. (a) $\omega_t=1.6238$, $D=2$ and $L=1.9347$, (b) $\omega_t=1.5856$, $D=2$ and $L=5.9405$.}}
\end{figure}

Results in Sec. \ref{sec3a} have shown that it is not at only one frequency that the total reflection of the same $k1$ and $k2$ modes can occur between the same two waveguides. It happens at different frequencies with different combinations of the two modes in amplitude and phase, and the phase change on reflection is not constant.
We can have some understanding of this phenomenon from modal scattering between two waveguides. In Fig. \ref{fig_12}, the number of propagative modes in the $+x$ direction in the narrow duct $N_{\rm{p,d}}$ is 1 or 2, thus one can find a combination of $k1$ and $k2$ modes in the wide duct (the cavity segment in Sec. \ref{sec3a}) that leads to zero transmission to the propagative channel(s) in the narrow duct (the mode coefficients of $k1$ and $k2$ are decided by the scattering matrix at the interface between the two waveguides). 
If $N_{\rm{p,d}}=3$, three incident waves ($k1$, $k2$ and $k3$) from the wide duct are needed to cancel each other in transmission to the propagative channels in the narrow duct. In Figs. \ref{fig_4} and \ref{fig_6} where $N_{\rm{p,d}}=2$, $k1$ and $k2$ modes and their left-travelling counterparts can cancel each other in transmission to the total four outgoing propagative channels of the two narrow ducts owing to the $x$-symmetry in the structure, which makes the interface scattering matrices at the left and right ends of the cavity identical. 
For $x$-symmetric structures like Figs. \ref{fig_2}, \ref{fig_4} and \ref{fig_6} but with $N_{\rm{p,d}}=3$ or for $x$-asymmetric structures with $N_{\rm{p,d}}=2$ in each of the two ducts, more than two coexisting resonant modes are then needed to underpin an embedded trapped mode, and to find it more tuning parameters are needed \citep{Hsu2016}. 
However, with $N_{\rm{p,d}}=1$ in the narrow ducts, we can still find embedded trapped modes characterized by two coexisting standing waves in $x$-asymmetric structures by tuning $L$ and $\omega$, as shown in Fig. \ref{fig_13}, since the total number of the outgoing propagative channels is 2. The interface scattering matrices at the two ends of the cavity segment are not identical (the two-by-two reflection matrices in Fig. \ref{fig_13}(a) are $\mathbf{R}_l(2\times 2)=[0.6706 + 0.0691\mathrm{i}, -0.1974 - 0.0381\mathrm{i}; -1.5578 - 0.3003\mathrm{i}, -0.0881 + 0.1879\mathrm{i}]$ and $\mathbf{R}_r(2\times 2)=[0.6706 + 0.0691\mathrm{i}, 0.1974 + 0.0381\mathrm{i}; 1.5578 + 0.3003\mathrm{i}, -0.0881 + 0.1879\mathrm{i}]$ respectively).

The wave trapping mechanism of the present embedded trapped modes associated with two-resonant-mode interference is essentially the same as that for the Fabry--P{\'e}rot type of trapped mode, in the sense of total reflection of travelling modes at the effective boundaries. It has been shown that in some cases \citep{Hein2012} a Fabry--P{\'e}rot trapped mode can be described by one standing wave between two resonators, which can totally reflect a single guided mode. Indeed, trapped modes or resonances in open systems can be understood as the result of total reflection of waves at effective boundaries. However, the component travelling waves can be rather complex and the effective boundaries are difficult to define in some cases, such as a high-lift configuration \citep{Duan2007}, which requires a sufficiently large computational domain with absorbing boundary conditions, and the mechanisms of total reflection can be different.

Total reflection for a particular combination of multiple travelling modes considered here is understood as a result of linear cancellation in transmission to the propagative channel(s) of the adjacent waveguide. It is noted that total reflection of a travelling acoustic mode or a mode-like wave can also occur owing to an acoustically hard wall, a resonator attached to a waveguide \citep{Rienstra2018}, the slowly varying cross-section of a duct that can support an acoustic turning point of the Wentze--Kramers--Brillouin (WKB) approximation where a slowly varying duct mode changes from cut-on to cut-off \citep{Rienstra2003}, the narrowing potential core of a jet flow \citep{Towne2017} and a shear layer in flow \citep{Martini2019,Rienstra2019,Saverna2019}.

\subsection{Fano scattering owing to two acoustic channels}
\label{sec3c}

\begin{figure}
	\begin{center}
		\includegraphics[width=13.5cm]{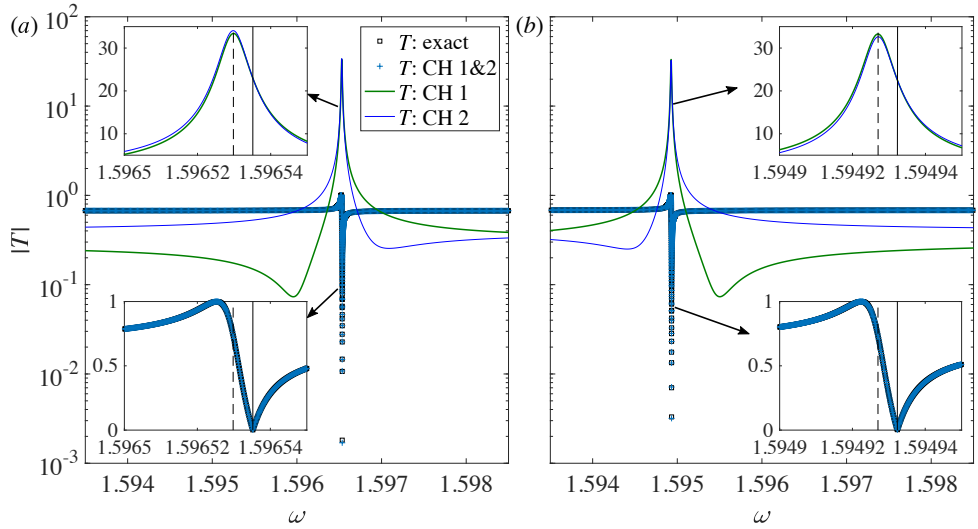}
	\end{center}	
	\caption{\label{fig_14} {(Colour online) Fano scattering phenomenon associated with the case in Fig. \ref{fig_2}. (a) $D=2$, $L=3.930$ and $\omega_{qt}=1.5965+4.9442 \times 10^{-6}\mathrm{i}$, (b) $D=2$, $L=3.945$ and $\omega_{qt}=1.5949+4.9516 \times 10^{-6}\mathrm{i}$. The vertical dashed lines denote $\mathrm{Re}(\omega_{qt})$ and the vertical solid lines denote the frequencies where $|T|=0$.}}
\end{figure}

As the cavity length is perturbed from the value for a trapped mode, the phase condition for total reflection at both effective boundaries is no longer fulfilled, then the trapped mode becomes quasi-trapped \citep{Hein2010,Hein2012}. Note that the symmetry-protected embedded trapped modes turn into quasi-trapped modes as the symmetry about the duct axis is broken \citep{Aslanyan2000,Pagneux2013}. For a quasi-trapped mode, the Fano scattering phenomenon \citep{Fano1961,Miroshnichenko2010,Lukyanchuk2010} can be observed from the numerical results of sound scattering by the cavity, as shown in Fig. \ref{fig_14}. To see the effects of the two resonant modes, the plane wave transmission from the left to the right duct is decomposed into three steps. The third step is the $k^+$ modes in the cavity exciting the plane wave in the right duct at $x=L$, with coefficients denoted by a column vector $\mathbf{T}_2$. The second step is the $k^+$ modes propagation in the cavity segment, denoted by a $N_2 \times N_2$ diagonal matrix $\mathbf{P}_d$ with on the diagonal the values of $\exp(-\mathrm{i} k^+_n L)$. The first step is the excitation of $k^+$ modes in the cavity by the incident wave, with the coefficients at $x=0$ denoted by a row vector $\mathbf{T}_1$. Then, the plane-wave transmission from the left to the right duct is formulated as, $T= \mathbf{T}_1 \mathbf{P}_d \mathbf{T}_2$. With this decomposition, one can observe the effects of multiple acoustic channels in the cavity on sound transmission. It is shown that the exact $|T|$ can be accurately represented by channel 1 plus channel 2, which are associated with $k1$ and $k2$ in the cavity. Two coexisting leaky resonances at the frequency of the quasi-trapped mode, i.e. $\mathrm{Re}(\omega_{qt})$, are observed. Moreover, $|T|=0$ happens not at $\mathrm{Re}(\omega_{qt})$ but at the frequency where $|T|$ associated with channels 1 and 2 cross, i.e. having the same amplitude and opposite phases. $|T|=1$ can be interpreted as the destructive-interference effect of the two leaky resonant modes on the reflection of the incident plane wave, leading to a zero reflection.

\section{Conclusion}

Acoustic embedded trapped modes associated with two-resonant-mode interference in two-dimensional duct--cavity structures have been calculated by the feedback-loop closure principle.
The full destructive interference in the radiation loss of two coexisting standing waves in a cavity open to two semi-infinite ducts has been numerically investigated. 
To the author’s best knowledge, for the first time the exact two coexisting resonant modes that underpin an embedded trapped mode have been demonstrated.

Results further indicate that the root cause of such a wave trapping characterized by two coexisting standing waves is total reflection of two guided waves at the interface between two waveguides. Such total reflection of multiple travelling modes is understood as a result of linear cancellation in transmission to the propagative channel(s) of the adjacent waveguide.
At the interface between the cavity segment and one of the two semi-infinite ducts, total reflection can occur for a particular combination of two propagative guided waves in the cavity. Total reflection requirements at both ends of the cavity segment decide the cavity length and frequency for a trapped mode, i.e. a perfect wave localization. 
It has been found in all the cases considered that the two standing waves are not two closed-cavity modes. 

For quasi-trapped modes, the Fano scattering phenomenon owing to the effects of two acoustic channels in the cavity on wave scattering has been shown.
$|T|=0$ and  $|T|=1$, owing to the full destructive interference of two coexisting leaky resonances in respectively transmission and reflection, happen near but not at the quasi-trapped mode frequency.

\section*{Appendix A. Convergence of calculations}

The convergence of calculations of trapped modes is presented on table \ref{tab:kd}. It is noted that with $N_1$ points in the $y$ direction in a 2-D duct, not all the $N_1$ duct modes solved are accurate. The solutions of some high-order evanescent modes are not accurate. However, errors for low-order propagative modes are small. With $N_1=400$ ($N_2=800$), the relative errors of $\exp(-\mathrm{i} k L)$ for $k1$ and $k2$ are smaller than $1.2 \times 10^{-4}$ in the trapped mode calculations, and the relative errors of $\exp(-\mathrm{i} k L_l)$ for $k1$ and $k2$ are smaller than $3.8 \times 10^{-4}$ in Fig. \ref{fig_12}, where $L_l$ denotes the plotted length of the left duct. The relative error is defined as $|\exp(-\mathrm{i} k L)-\exp(-\mathrm{i} k_a L)|/|\exp(-\mathrm{i} k_a L)|$, where $k_a$ is the analytical solution. The converged results of trapped modes can be achieved, because the high-order transverse modes decay quickly in propagation so that they play an unimportant role in the trapped modes presented which are constructed by a few low-order propagative transverse modes, and a rather large number of points in the transverse direction are used. It is noted that, for the same duct--cavity structures, the converged results of trapped modes at high frequencies require more discrete points, which is similar to the global eigenvalue approach \citep{Hein2004,Koch2005,Duan2007,Hein2007,Hein2010,Hein2012}.

\begin{table}[hbt!]
\caption{\label{tab:kd} Convergence of calculations of trapped modes}
\centering
\begin{tabular}{lccc}
\hline
   Case &~~    $N_1$ &~~   $\omega_t$   &~~    $L$  \\ [3pt]
\hline
   Fig. \ref{fig_6} &~~   400  &~~  5.39709097 & ~~2.43562979\\
   Fig. \ref{fig_6} &~~   600  &~~  5.39714028 & ~~2.43560621 \\
   Fig. \ref{fig_6} &~~   800  &~~  5.39713748 & ~~2.43561051 \\
   Fig. \ref{fig_10} &~~   400  &~~   1.68525467 & ~~11.1849953\\
   Fig. \ref{fig_10} &~~   600  &~~   1.68525876 & ~~11.1849626\\
   Fig. \ref{fig_10} &~~   800  &~~   1.68525859 & ~~11.1849747\\
\hline
\end{tabular}
\end{table}



\end{document}